\documentclass[reprint,amsmath,amssymb,aip,jcp,author-numerical]{revtex4-1}
\usepackage{graphics}
\usepackage{graphicx}
\usepackage{dcolumn}
\usepackage{bm}
\usepackage{natbib}
\usepackage{hyperref}
\usepackage{color}

\usepackage{tikz}

\begin{document}
	
	\title{Electrostatic interactions in water: nonlocal electrostatic approach}
	\author{M. Vatin}
	\affiliation{Sorbonne Universit{\'e}, CNRS, Laboratoire de Physique Th{\'e}orique de la Mati{\`e}re Condens{\'e}e (LPTMC, UMR 7600), F-75005 Paris, France}
	\affiliation{ICSM/LMCT Site de Marcoule, F-30207 Bagnols-sur-Cèze Cedex France}
	\author{A. Porro}
	\affiliation{Sorbonne Universit{\'e}, CNRS, Laboratoire de Physique Th{\'e}orique de la Mati{\`e}re Condens{\'e}e (LPTMC, UMR 7600), F-75005 Paris, France}
	\author{N. Sator}
	\affiliation{Sorbonne Universit{\'e}, CNRS, Laboratoire de Physique Th{\'e}orique de la Mati{\`e}re Condens{\'e}e (LPTMC, UMR 7600), F-75005 Paris, France}
	\author{J-F Dufr\^eche}
	\affiliation{ICSM/LMCT Site de Marcoule, F-30207 Bagnols-sur-Cèze Cedex France}
	\author{H. Berthoumieux}
	\affiliation{Sorbonne Universit{\'e}, CNRS, Laboratoire de Physique Th{\'e}orique de la Mati{\`e}re Condens{\'e}e (LPTMC, UMR 7600), F-75005 Paris, France}
	\begin{abstract}
		Can we avoid molecular dynamics simulations to estimate the electrostatic interaction between charged objects separated by a nanometric distance in water? To answer this question, we develop a continuous model for the dielectric properties of water based on a functional of the polarization.  A phenomenological Landau-Ginzburg Hamiltonian for the electrostatic energy of water is parameterized to capture the dipolar correlations in the fluid at the nanometric scale.
		We show that in this framework, the effective interactions of simple objects such as point charges are analytically tractable. In particular, the derivation of the interaction energy between a solvated charge and a surface can be reduced to a system of linear equations of electrostatic potentials and analytically solved.
		This approach could thus give access in few calculation lines to data that necessitate long and costly simulations.
	\end{abstract}
	\maketitle

	\section{Introduction} 
	It is difficult to overstate the importance of electrostatic interactions in aqueous solutions in nanometric confinements: interaction of charged solutes with proteins in a biological cell, ions in nanoporous materials...\cite{jubin2018} The description of the solvent in these systems is challenging \cite{carof2014} as one can not skip the molecular details of water at this scale but running a all-atoms simulation as a prerequisite of the study of such systems can appear vainly costly in terms of computer time. 
	
	The dielectric interaction between ions can be obtained by modeling water as a dielectric medium characterized by its permittivity. If one represents the ions as point charges and water as a local dielectric medium characterized by the bulk dielectric relative permittivity $\epsilon_w=78.3$ at 25~$^\circ$C, the problem is a textbook exercise of electrostatics\cite{jackson1999classical}. The analytic result obtained for monovalent ions is surprisingly a very good estimation of the molecular dynamics results for an interaction distance larger than 1 or 2 nm. But at small distance, the interaction depends on the short-range molecular structure of the solvent which is not included in this macroscopic description of the fluid. 
	
	Water is an associated liquid, structured by a network of intermolecular interactions, the hydrogen bonds, which induces short range correlations of the molecular orientations. The bond charge structure factor $S_c(q)=\langle \rho (q) \rho(-q) \rangle$, were $\rho(q)$ is the bond charge density in Fourier space, and its associated response function, the  longitudinal dielectric susceptibility $\chi_{\parallel}(q)$ were consistently determined from molecular dynamics simulations and experimental measurements at the end of the nineties \cite{bopp1996}. In the Fourier space, $\chi_{\parallel}(q)$ possesses a pronounced maximum, far above 1, for $q=3$ \AA$^{-1}$ which is the signature of an overscreening effect of the solvent, {\it i. e.} the possibility for two charges of the same sign to attract each other at short distance\cite{kornyshev1997}, and reveals the existence of characteristic lengths of molecular scale for the bond charge structure. More recently the dipole-dipole correlation function was obtained with MD simulations of SPC/E water \cite{zhang2014}. At short range, the correlation function between two dipoles oscillates and exponentially decays and is enhanced when compared to long-range van der Waals dipole-dipole correlations that are reached for separations larger than 1.5 nm.

	The theory of nonlocal electrostatics makes possible to include short-range correlation lengths between dipoles of solvent molecules in a continuous description.  This framework is based on a Landau-Ginsburg approach to express the electrostatic energy of the medium as a functional of the polarization\cite{maggs2006,paillusson2010,berthoumieux2015fluctuation,berthoumieux2018}. Increasing the complexity of this functional generates a zoology of $\chi_{\parallel}(q)$ and it is possible to propose a model associated with an overscreening response.  
	These phenomenological models have been used to study the effect of the non locality on the Born solvation energy of ions in water, with an analytic approach for a simple nonlocal functional \cite{hildebrandt2004novel} and numerically for a model including the overscreening effects \cite{rottler2009}. The sign and the amplitude of the correction to the local Born solvation energy depends on the model.
	
	However, the nonlocal dielectric models proposed until now strongly overestimate the range of the dipolar correlations in water and thus poorly reproduce the dielectric properties of bulk water \cite{fedorov2007,berthoumieux2015fluctuation}.  Such models were parameterized using the longitudinal dielectric susceptibility of SPC/E water \cite{jeanmairet2013} in which a molecule is composed of three partial charges, two positive for the hydrogen and one negative for the oxygen (see Fig.1) and its charge distribution  can thus be decomposed in a dipole, quadrupole etc... The charge correlations of liquid water are the sum of the self and cross correlations of all the multipoles of the molecule. Such a complexity can difficultly be included  in a model based on a polar field and containing few parameters. Non-Gaussian functional that include saturation effects have been envisaged to improve the description of electrostatic in water \cite{kornyshevsat1997,levy2012,paillusson19}. However it is not clear why nonlinear effects should be necessary to describe dipolar correlations in pure water.
	
	In this work, we propose to parameterize the electrostatic functional, which is the input of the nonlocal dielectric framework, with the dielectric susceptibility of SPC/E water treated as a dipolar fluid. This consists in replacing the three charge sites in the SPC/E explicit model of water in MD  by a two charge distribution associated with the same dipolar moment. We show that this approach  allows the properties of bulk water such as polar correlation functions to be reproduced with good accuracy. In a second time, we use the field theory framework to derive analytically the electrostatic interactions between point charges and dipoles. 
	Finally we consider the interaction between a point charge and a surface and illustrate using this example how to calculate electrostatic interactions between extended objects and implement the boundary conditions in this framework. The last part is devoted to the conclusion.
	
	\section{Water as a polar fluid}
	We propose to describe water as a polar fluid using the polarization field ${\bf P}(r)$ as a relevant physical parameter to express the  electrostatic energy as \cite{maggs2006, berthoumieux2018,berthoumieux2015fluctuation},
	\begin{eqnarray}
	\label{Hchi}
	\mathcal{H}[\mathbf{P}]&=&\frac{1}{2\epsilon_0} \int d^3r d^3r'\mathbf{P_r}\cdot\bm{\chi}_{r,r'}^{-1}\cdot\mathbf{P_r'}  \\
	&=& \frac{1}{2\epsilon_0} \int d^3r d^3r'\mathbf{P_r}\cdot\bm{K}_{r,r'}\cdot\mathbf{P_r'}\nonumber\\
	&+& \frac{1}{2\epsilon_0} \int d^3r d^3r' \frac{\nabla \cdot {\bf P}(r) \nabla \cdot {\bf P}(r')}{4\pi| r-r'|}.
	\label{Hq}
	\mbox{}\end{eqnarray}
	where $\epsilon_0$ is the vacuum permittivity and $\bm{\chi}_{r,r'}$ the two-point dielectric susceptibility tensor. This electrostatic energy can be decomposed into a term coming from the  short-range molecular interactions encoded by the kernel  $\bm{K}_{r,r'}$ and a long range Coulombic interaction. For isotropic systems, the susceptibility can be decomposed in a longitudinal $\chi_{\parallel}$ and in a transverse $\chi_{\perp}$ part that are linked to the kernel  $\bm{K}$ in the Fourier space by the relations
	\begin{eqnarray}
	\label{chilongtrans}
	\chi_{ij}(q)&=&\chi_{\parallel}(q)q_iq_j/q^2+\chi_{\perp}(q)\left(\delta_{ij}-q_iq_j/q^2\right),\\
	\label{chiparallel}
	\chi_{\parallel}(q)&=&\left(1+K_{\parallel}(q)\right)^{-1}, \quad \chi_{\perp}(q)=K^{-1}_{\perp}(q),
	\end{eqnarray}
with $(i,j=x,y,z)$. 
	Following a Landau-Ginsburg approach, one can propose a local expression for the kernel {\bf K} that will give rise to desired properties \cite{maggs2006}.
	
	Molecular dynamics simulations have been used to determine the susceptibility ${\bf \chi}(q)$ of explicit models of water such as SPC/E model by evaluating polarization correlations that can be linked to the dielectric susceptibility via the fluctuation dissipation theorem,
	\begin{equation}
	\label{chiparallelTFD}
	\chi_{\parallel}(q)=\frac{\beta}{\epsilon_0}\langle\frac{{\bf q}\cdot{\bf P}({\bf q}){\bf q}\cdot{\bf P}(-{\bf q})}{q^2}\rangle.
	\end{equation} 
	However, the nonlocal dielectric models \cite{berthoumieux2015fluctuation,berthoumieux2018} parameterized on this response function reproduce poorly the properties of bulk water as it overestimates by a factor 3 or 4 the range of the dipolar correlations in the bulk water.
	
	In order to improve the parametrization of the field model, we propose to project the SPC/E model of water onto a model of dipolar symmetry, that can be captured by the continuous model which is developed using the polarization field $P(r)$ and can reproduce only the behavior of water at the dipolar order.  To do so, we replace in each equilibrium configuration of SPC/E water obtained with molecular dynamics the 3 charge water molecules by a 2 charges distribution associated with a dipolar moment, $\mu$=2.35 D of same value and direction as the one of SPC/E water molecules. The oxygen is treated as usual in the SPC/E model, {\it i.e.} an atom  of charge $q_0=-0.8476$ $ e $ and associated with a Lennard-Jones center.  A dummy atom, X, associated with a charge $q_f=0.42$ $e$ is placed on the bisector of the angle HOH at a distance $d=1$ \AA \quad of the oxygen. This is illustrated in Fig. 1.
	\begin{figure}
		\includegraphics[scale=0.8]{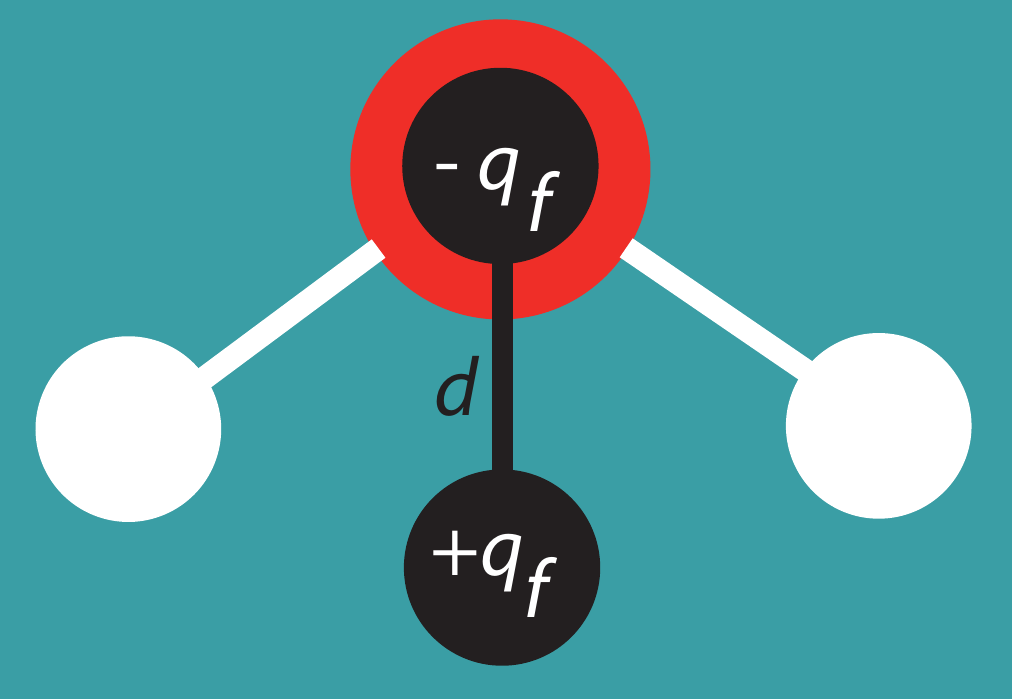}
		\caption{Sketch of a water molecule and of its representation as a dipolar molecule of same dipole moment. The charge $q_f$ is taken equal to $q_f=0.42$e  and $d=1 $ \AA }
	\end{figure}
	This simple two-charge model is named \textit{Dipolar Dumbbell} model (\textit{DD})\cite{raineri92}.
	We calculate the susceptibility of the corresponding fluid following a protocol detailed in Appendix B. 
	
	\begin{figure}
		\includegraphics[scale=0.22]{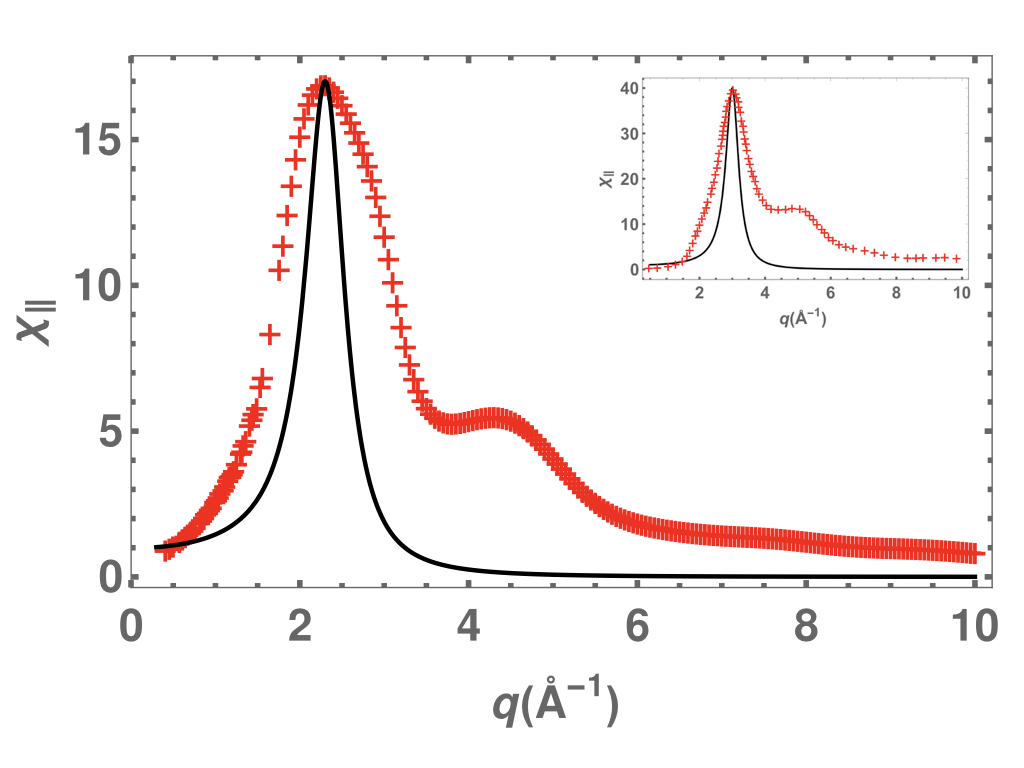}
		\caption{Longitudinal susceptibility derived with Molecular dynamics simulations and Field theory model. The susceptibility plotted in red is obtained using MD  of SPC/E water projected on a two point {\it DD} model. The susceptibility obtained from the Field theory framework is given by Eq. (\ref{chil}) for $K=1/70$, $\kappa_l=-0.36$ \AA$^2$, $\alpha=0.034$ \AA$^4$. The inset represents the longitudinal susceptibility for the SPC/E model (red curve) and the associated field theory susceptibility (black curve) obtained for parameters K=1/70, $\kappa_l=-0.22$\AA$^2$, $\alpha=0.012$\AA$^4$ corresponding to a reproduction of the SPC/E susceptibility}
	\end{figure}
	The shape of the longitudinal susceptibility in Fourier space of the water DD model, represented in Fig. 2, is similar to the susceptibility of SPC/E model (represented in the inset of the Fig. 2), as it presents a pronounced maximum, signature of the over-screening in water and possesses the same bulk susceptibility, obtained for $q=0$. The position and value of the maximum differ however.
	
	We propose the following Hamiltonian, 
	\begin{eqnarray}
	& &\mathcal{H}[\mathbf{P}]=\frac{1}{2\epsilon_0} \int d^3r(K\mathbf{P(r)}^2+\kappa_l(\mathbf{\bm{\bm{\nabla}}}\cdot\mathbf{P(r)})^2\nonumber\\ &+&\alpha\left(\mathbf{\bm{\bm{\nabla}}}^2\cdot \mathbf{P(r)}\right)^2
	+\frac{1}{2\epsilon_0} \int d^3rd^3r'\frac{\mathbf{\bm{\bm{\nabla}}}\cdot\mathbf{P(r)}\mathbf{\bm{\bm{\nabla}}}\cdot\mathbf{P(r')}}{4\pi|\mathbf{r-r'}|}
	\mbox{.}
	\label{eq:H}\end{eqnarray}
	associated with the longitudinal susceptibility $\chi_{\parallel}(q)$ 
	\begin{eqnarray}
	\label{chil}
	\chi_{\parallel}(q)&=&\frac{1}{1+K+\kappa_l q^2+\alpha q^4}
	\end{eqnarray}
	determined using Eqs. (\ref{Hchi}, \ref{chilongtrans}) to model the susceptibility obtained using MD simulations and presented in Fig. 2. Note that the transverse susceptibility is in this case equal to $\chi_{\perp}(q)=1/K$.
	Such a function presents a quasi-resonant behavior for well-chosen values of $\kappa_l$ and $\alpha$ \cite{berthoumieux2018}.

	The longitudinal susceptibility given in Eq. (\ref{chil}) is characterized by three parameters $(\alpha, \kappa_l, K)$ and we fix their values to reproduce nonlocal dielectric properties of the $DD$ model of water:  first, the bulk dielectric susceptibility, $\chi_{\parallel}(q=0)=\chi_b$, the characteristic of the quasi-resonant maximum, {\it i. e.} its position, $q_0$ and its value $\chi_m$. 
	We extract the values $\chi_b=1.01$, $q_0=2.3$ \AA$^{-1}$ and $\chi_m=17$) from the simulations and we fix the values of the parameters of the continuous model $K$, $\kappa_l$ and $\alpha$ through the relations,
	\begin{eqnarray}
	\chi_b=\frac{1}{1+K},
	\quad  
q_{0}^2	=\frac{-\kappa_l}{2\alpha},
	\quad
\frac{1}{\chi_m}=1+K-\frac{\kappa_l^2}{4\alpha}.
	\label{param3}
	\end{eqnarray}
 It gives $K$=1/70, $\kappa_l$=-0.36 \AA$^{2}$ and $\alpha$=0.034 \AA$^{4}$.  The corresponding field theory susceptibility, obtained by parameterizing Eq. (\ref{chil}) using the SPC/E projection on the $DD$ model is plotted in black in Fig. 2. The inset represents the Field theory susceptibility parameterized directly on SPC/E model. 
	
	In order to validate the continuous model proposed here for the description of the dielectric properties of bulk water, we compare the polarization correlation $\langle P(0)P(r)\rangle_{\parallel}$ derived analytically in the first framework and numerically using molecular dynamics simulations.

	The correlations are linked to the susceptibility response in the real space  $\chi_{ij}(r)$ via the fluctuation-dissipation theorem that can be written as
	\begin{equation}
	\label{PP}
	\langle P(0)P(r) \rangle_{i,j}=\epsilon_0k_BT\chi_{ij}(r).
	\end{equation}

	The susceptibility tensor $\chi_{ij}(r)$ is obtained by calculating the Fourier transform of $\chi_{ij}(q)$ . 
	The longitudinal correlation of the polarization obeys,
	\begin{equation}
	\label{longcorP}
	\langle P(0)P(r) \rangle_{\parallel}=\epsilon_0k_BT I_2(r),
	\end{equation}
	with
	\begin{eqnarray}
	\label{I2}
	I_2(r)&=&\frac{1}{(2\pi)^3}\int d^3qe^{i{\bf q}\cdot{\bf r}}\chi_{zz}(q)\nonumber\\
	& =&\frac{1}{(2\pi)^3}\int d^3qe^{i{\bf q}\cdot{\bf r}}\Bigg(\left(\frac{1}{1+K+\kappa_l q^2+\alpha q^4}-\frac{1}{K}\right)\frac{q_z^2}{q^2}\nonumber\\
	&+&\frac{1}{K}\Bigg)
	\end{eqnarray}
	
	Performing the Fourier transform of Eq. (\ref{I2}), we obtain for the longitudinal correlation function
	\begin{widetext}
		\begin{eqnarray}
		\label{correlfunction}
		\langle P(0)P(r) \rangle_{\parallel}(r)&=&\frac{\epsilon_0k_BT}{2\pi K(K+1)r^3}
		-\frac{\epsilon_0 k_BT e^{-r/\lambda_e}}{4 \pi (K+1) r}{\rm cos}\left(r/\lambda_o\right)\left(\frac{2}{r^2}+\frac{1}{r\lambda_o}\left(R+\frac{1}{R}\right)\right)\nonumber\\&-&\frac{\epsilon_0k_BT e^{-r/\lambda_e}}{4 \pi (K+1) r}{\rm sin}(r/\lambda_o)\Big(\frac{1}{\lambda_o^2}\left(\frac{1}{2R^3}+\frac{1}{R}+\frac{R}{2}\right) +\frac{1}{r}\left(\frac{1}{\lambda_o}+\frac{1}{\lambda_o R^2}\right)+\frac{1}{r^2}\left(\frac{1}{R}-1\right)\Big)
		\end{eqnarray} 
	\end{widetext}
	with
	\begin{equation}
	\label{lambda}
	\lambda_e=\frac{\sqrt{2}}{q_0\sqrt{1/\sqrt{\zeta}-1}}, \quad \lambda_o=\frac{\sqrt{2}}{q_0\sqrt{1/\sqrt{\zeta}+1}},
	\end{equation}
	where $q_0^2=-\kappa_l/2\alpha$, $\zeta=\alpha q_0^4 \chi_b$ and $R=\lambda_e/\lambda_o$. For the chosen set of parameters, one finds $\lambda_e=$3.8 \AA \quad and $2\pi \lambda_o$=2.7 \AA. The polarization correlation is the sum of a long-range function, $\epsilon_0k_BT/2\pi K(K+1)r^3$ that depends only on the bulk properties of the fluid, and of a nonlocal term taking the form of an oscillating function in a decaying envelope dominant at small range and negligible over a nanometric distance.
	
	The dipole-dipole correlations obtained using the field theory framework,
	\begin{equation}
	\label{mumuFT}
	\langle \mu(0)\mu(r) \rangle_{FT}=\langle P(0)P(r) \rangle_{\parallel}/\mu_D^2\rho_0^2,
	\end{equation}
		 can be compared  to water dipole correlations $\langle \mu(0)\mu(r) \rangle_{MD}$ obtained from molecular dynamics simulations\cite{zhang2014}. 
	
	The expression given in Eq. (\ref{mumuFT}) is represented in Fig.2 (black curve) for the parameters values obtained using Eq. (\ref{param3}) and compared to the longitudinal dipolar correlation obtained with molecular dynamics simulations for SPC/E model (red cross). The black curve in the  inset represents the dipolar correlation given in Eq. (\ref{mumuFT}) obtained for a functional parameterized with the longitudinal susceptibility of SPC/E water \cite{berthoumieux2015}.  As one sees in this case, the model strongly overestimates the correlations and does not reproduce the water layering around a reference molecule, the extrema of the correlation functions being shifted with the MD data. On the contrary, the continuous model parameterized using only the dipolar correlations in SPC/E liquid reproduces the range and the structure of the nonlocal correlations.  
	\begin{figure}
		\includegraphics[scale=0.45]{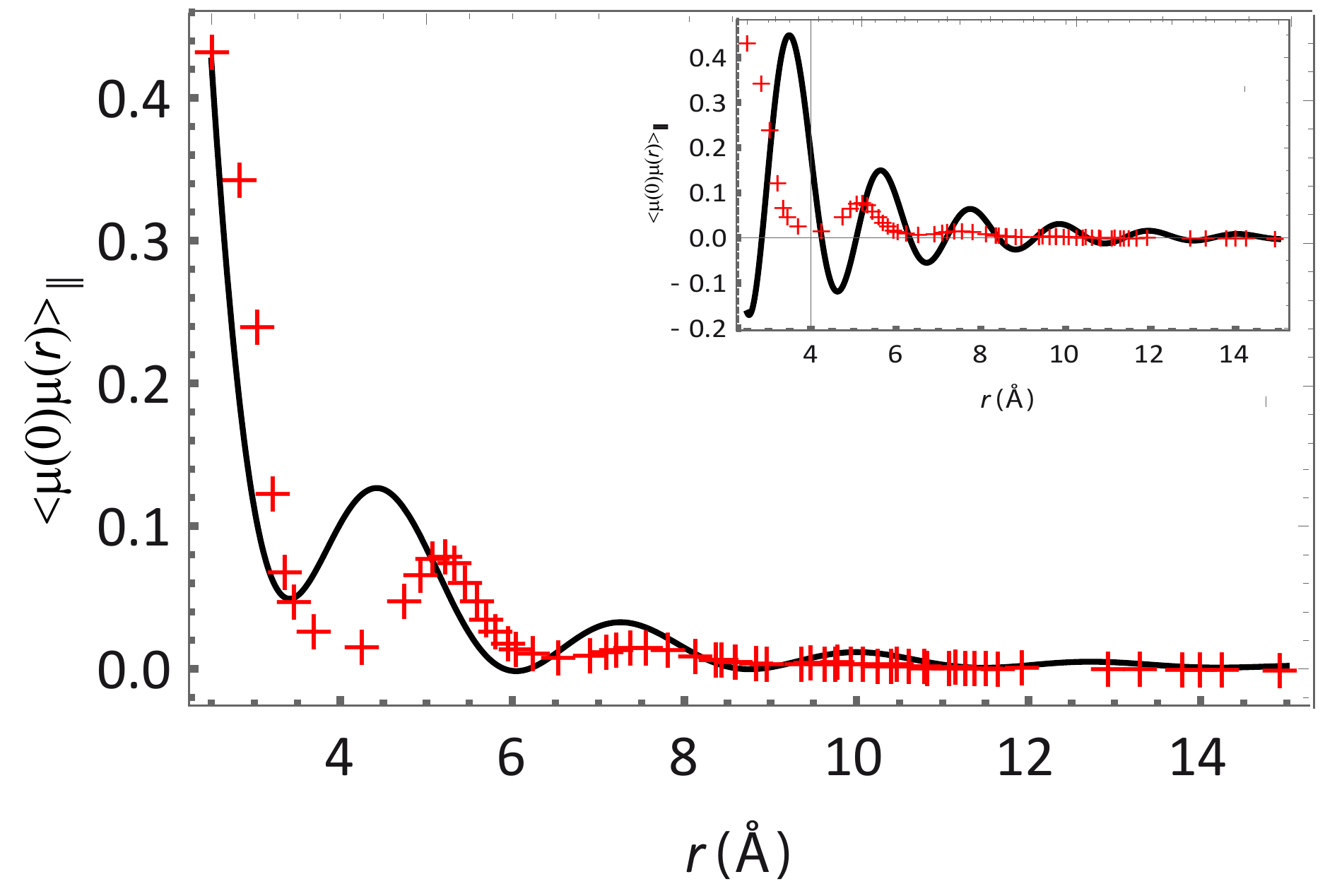}
		\caption{Dipolar correlation function. The red cross are results from molecular dynamics simulations that are reproduced from the reference\cite{zhang2014}. The black curve is plotted using the expression $\langle \mu(0)\mu(r) \rangle_{FT}$ given in Eq. (\ref{mumuFT}) and the parameters given in Fig. 2. The inset represents the dipolar correlation obtained for parameters K=1/70, $\kappa_l=-0.22$\AA$^2$, $\alpha=0.012$\AA$^4$ corresponding to a reproduction of the SPC/E susceptibility.}
	\end{figure}

	\section{Interaction between two point charges and between two dipoles }
	In this section, we first calculate the electrostatic interaction between two point charges  $Q_\alpha$, with $\alpha=1,2$ immersed in a nonlocal dielectric medium, located in ${\bf r}_\alpha$, and separated by a distance $r=|{\bf r}_1-{\bf r}_2|$ (see Fig. 4a). 
	The electrostatic potential $\phi(r)$ generated by a point charge $Q_\alpha$ located at the position  $\bf{r}=\bf{r_\alpha}$ can be written as, 
	\begin{eqnarray}
	\phi({\bf r}-{\bf r}_\alpha)=\frac{Q_\alpha}{\epsilon_0}\int d^3 q \frac{e^{-i\mathbf{q\cdot(r-r_i)}}}{q_{i} \epsilon_{ij}(\mathbf{q}) q_{j}}
	\label{phi}
	\mbox{,}\end{eqnarray}
	where the dielectric permittivity tensor in Fourier space $\epsilon_{ij}(\mathbf{q})$,  $(i, j= x,y,z)$  is linked to the susceptibility as follows \cite{kornyshev1997} :
	\begin{eqnarray}
	{\epsilon}_{ij}(\bm{q})&=&\left({\bf I}-{\bm \chi}({\bm q})\right)^{-1}_{ij}\nonumber\\
	&=&\epsilon_{\parallel}(q)\frac{q_iq_j}{q^2}+\epsilon_{\perp}(q)\left(\delta_{ij}-\frac{q_iq_j}{q^2}\right),
	\label{eq:chiepsilon}
	\end{eqnarray}
	with ${\bf I}$, the identity matrix and the longitudinal and transverse parts can be written as
	\begin{eqnarray}
	\epsilon_{\parallel}(q)=\frac{1}{1-\chi_{\parallel}(q)},  \quad \epsilon_{\perp}(q)=\frac{1}{1-\chi_{\perp}(q)} .\end{eqnarray}
	
	The interaction energy between two point charges $Q_1$ and $Q_2$  and located in $r=r_1$, respectively $r=r_2$ is 
	\begin{eqnarray}
	U_{nl}(\mathbf{r_1-r_2})=\frac{Q_1Q_2}{(2\pi)^3\epsilon_0}\int d^3q\frac{e^{-i\mathbf{q\cdot(r_1-r_2)}}}{q^2\epsilon_{\parallel}(q)}
	\label{eq:U}
	\mbox{.}\end{eqnarray}
	Performing the integral using the residue theorem, one finds for the interaction energy $U_{nl}(\mathbf{r_1-r_2})$ 
	\begin{eqnarray}
	\label{Unl}
	U_{nl}(r)=\frac{Q_1Q_2}{4 \pi  r \epsilon_0 \epsilon_w}\Bigg(1&+&\frac{(\epsilon_w-1) e^{-\frac{r}{\lambda_e}} }{2 \lambda_e \lambda_o}\Big(\left(\lambda_o^2-\lambda_e^2\right) \sin \left(\frac{r}{\lambda_o}\right)\nonumber\\
	&+&  2 \lambda_e \lambda_o \cos \left(\frac{r}{\lambda_o}\right)\Big)\Bigg)
	\label{eq:chch}
	\mbox{.}\end{eqnarray}
	with $r=|{\bf r}_1-{\bf r}_2|$ and with $\epsilon_w=1/(1-\chi_b)$ is the macroscopic susceptibility.
	As the polarization correlation function given in Eq. (\ref{correlfunction}), it contains a long-range local term $U_{loc}(r)=\frac{Q_1Q_2}{4 \pi  \epsilon_0 \epsilon_w r}$ that depends only on the bulk properties of the solvent and an oscillating decaying contribution due to the fine structure of the solvent.

	We now consider the interaction between two dipoles, each composed of two point charges of opposing sign ($q$, -$q$) and separated by a small distance $l$. The geometrical center of the dipole 1, dipole 2 respectively, is located in $r_1=0$, $r_2=r$ respectively.  The positive charge of the dipole 1, of the dipole 2 respectively, is located in ${\bf r}={\bf r}_1^+$, respectively ${\bf r}={\bf r}_2^+$, and the negative charges in ${\bf r}=r_1^-$ and ${\bf r}=r_2^-$: 
	\begin{widetext}
		\begin{eqnarray} 
		\mathbf{r_1^{+}}&=&\left(0,\frac{l}{2}  \cos (\theta_1),\frac{l}{2}  \sin (\theta_1)\right) \quad
		\mathbf{r_2^{+}}=\left(\frac{l}{2}  \cos (\theta_2) \cos (\phi ),\frac{l}{2}  \cos (\theta_2)+r,\frac{l}{2}  \sin (\theta_2) \cos (\phi )\right),\nonumber \\ \mathbf{r_1^{-}}&=&-\mathbf{r_1^{+}}, \quad
		\mathbf{r_2^{-}}=\left(-\frac{l}{2}  \sin (\theta_2) \cos (\phi ),-\frac{l}{2}  \cos (\theta_2)+r,-\frac{l}{2}  \sin (\theta_2) \cos (\phi )\right)
		\mbox{.}
		\end{eqnarray} 
	\end{widetext}
	The angles $\theta_i$, ($i=1,2$) vary between 0 and $\pi$, and $\phi$ varies between 0 and $2\pi$. They  are defined in Fig.2 b. 
	The medium being linear, the interaction energy between the dipoles is the sum of the pair interaction between the point charges of dipole 1 and dipole 2, as
	\begin{widetext}
	\begin{eqnarray} 
	\label{dipolE}
	\mathcal{W}_{nl}(r,\theta_1 ,\theta_2,\phi)=U_{nl}(|\mathbf{r_1^{+}-r_2^{+}}|)+U_{nl}(|\mathbf{r_1^{-}-r_2^{-}}|)-U_{nl}(|\mathbf{r_1^{-}-r_2^{+}}|)-U_{nl}(|\mathbf{r_1^{+}-r_2^{-}}|).
	\end{eqnarray}
	\end{widetext}
	where $U_{nl}(r)$ is the charge-charge interaction given in Eq.(\ref{eq:U}).
	Assuming $l/r\ll1$ and expanding each term of Eq.(\ref{dipolE}) around $r$, we find
	\begin{widetext}
		\begin{eqnarray} 
		\label{Wnl}
		\mathcal{W}_{nl}(r,\theta_1 ,\theta_2,\phi)	&=&-l^2 \Bigg(\cos (\theta_1) \cos (\theta_2) U_{nl}''(r)+\frac{\sin (\theta_1) \sin (\theta_2) \cos (\phi ) U_{nl}'(r)}{r}\Bigg)+O(l^2)
		\mbox{,}\end{eqnarray}
	\end{widetext}
	The free interaction energy $\mathcal{W}(r)$ between the dipoles separated by a distance $r$ is obtained by taken into account all the possible orientations of the dipoles and by assuming a Boltzmann statistic for each orientation,

	\begin{eqnarray}
	\label{Wnlaverage}
	\mathcal{W}_{nl}(r)=-k_B T \ln(\mathcal{Z}_{nl}(r))
	\end{eqnarray}
	with
\begin{widetext}
	\begin{eqnarray}
	\label{Z}
	\mathcal{Z}_{nl}(r)=\frac{1}{8 \pi} \int_0^{2 \pi} d\phi \int_0^{\pi} \sin (\theta_1) d \theta_1 \int_0^{\pi} d \theta_2  \sin (\theta_2) e^{-\beta\hphantom{.} \mathcal{W}_{nl}(r,\theta_1 ,\theta_2,\phi)}.
	\end{eqnarray}
	To estimate the free interaction energy, we first perform analytically the integration over $\phi$ in Eq. (\ref{Z}) and obtain,
		\begin{eqnarray}
		Z_{nl}(r)= \frac{1}{8 \pi}\int_0^{ \pi} \sin(\theta_1)d\theta_1 \int_0^{ \pi} \sin(\theta_2)d\theta_2 \hphantom{.} e^{-\beta l^2 \cos (\theta_1) \cos (\theta_2) U_{nl}''(r)} I_0\left(\frac{l^2 \beta  \sin (\theta_1) \sin (\theta_2) U_{nl}'(r)}{r}\right)
		\mbox{,}\end{eqnarray}
	\end{widetext}
	where $I_0(x)$ is the modified Bessel function of the first order. 
	The dipolar free interaction energy $\mathcal{W}_{nl}(r)$, given in Eq. (\ref{Wnlaverage}), is  
	obtained by computing numerically the integration over $\theta_1$ and $\theta_2$ for $\mathcal{Z}_{nl}(r)$.
	Note that for a local dielectric description of water, the interaction energy $\mathcal{W}_{loc}(r,\theta_1 ,\theta_2,\phi)$, obtained by replacing $U_{nl}(r)$ by $U_{loc}(r)$ in Eqs. (\ref{dipolE},\ref{Wnl}), is much smaller than $k_BT$ at room temperature. The corresponding partition function can be expanded at the first order in $\mathcal{W}_{loc}/k_bT$ and one obtains an analytic expression for the Keesom (permanent-permanent dipoles) interaction energy as \cite{israelachvili1991intermolecular}
	\begin{equation}
	\label{Wloc}
	\mathcal{W}_{loc}(r)=\frac{-(l q)^4}{3(4\pi\epsilon_0\epsilon_w)k_BTr^6}.
	\end{equation} 
	
	\begin{figure}
		\includegraphics[scale=0.7]{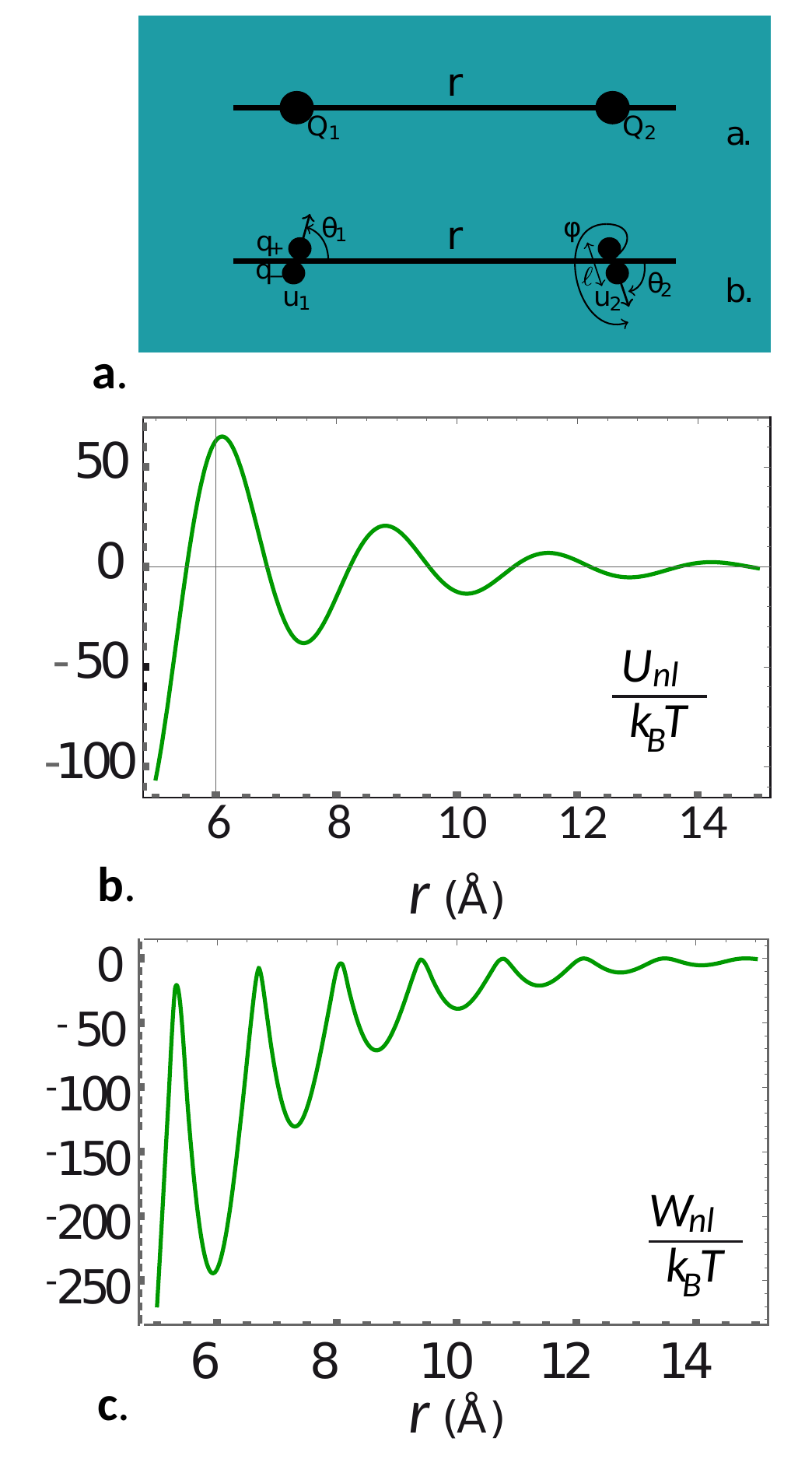}
		\caption{Electrostatic interaction in a nonlocal medium. {\bf a.} Sketch of the ion-ion and dipole-dipole interactions. {\bf b.} Interaction energy of two point charges with $Q_1$=-$Q_2$ =1.6$\times$10$^{-19}$ C  rescaled by $k_bT$. The expression is given in Eq. (\ref{Unl})  and the parameters used are given in Fig.2. {\bf c.} Interaction energy between two freely rotating dipoles ( $q_+$=-$q_-$=1.6 $\times$10$^{-19}$ C rescaled by $k_bT$. The expression is given in Eq. (\ref{Wnlaverage}) and the parameters used are given in Fig.2.}
	\end{figure}
	
The interaction energy between two point charges $U_{nl}(r)$ and the interaction energy between freely rotating dipoles are plotted in Fig. 4 {\bf b.} and Fig. 4 {\bf c.} respectively.   $U_{nl}(r)$ can be compared to the potential of mean force at infinite dilution (McMillan-Mayer potential) between two small atomic ions calculated from molecular dynamics simulation \cite{lyubartsev1995,Molina2011,Kalcher09}. Qualitatively an oscillating potential around a Coulomb law whose period is around 2~\AA \ is recovered.  The interactions are not monotonous at short distance. $U_{nl}(r)$ presents a succession of extrema corresponding to the layering of the solvent. Nevertheless, the agreement is far from being quantitative because the amplitudes of the oscillations are much larger ($\approx$ 50 $k_BT$ at short distance) than the ones observed in the interaction potential between small ions determined through MD simulations ($\approx k_B$T at the same distance). The point charges used here give rise to much stronger electrostatic fields than the ones generated by the charged Lennard-Jones spheres used in MD to describe ions. The diameter of the ions strongly reduces the local and non local polarization of the solvent. The next step of this work is to derive interactions between Smeared Born Spheres \cite{kornyshev1996} that reproduce better the properties of ions in water.

A similar trend is observed for $\mathcal{W}_{nl}(r)$ which presents successive minima with a period of 2 \AA. Such intense oscillations are not observed in the interaction potential of two freely rotating ion pairs that can be obtained from Molecular Dynamics \cite{Molina2011}. The difference can also be related to the hard repulsion of the solute.

	\section{Interaction of a point charge with a surface}
	The interaction of a charge solvated in water with a surface of a dielectric is of prime importance in many processes such as electrochemistry, biochemistry etc... In this section we evaluate the interaction energy between a point charge solvated in a nonlocal dielectric medium and a  dielectric medium of permittivity $\epsilon_d$. 
	
	An infinite flat surface separating the local and the nonlocal media is located in $z=0$. A charge $q$ is placed at a distance $d$ of the surface in the nonlocal medium and its coordinates can be noted by $(0,0,d)$ in a cylindrical referential (see Fig. 5).  As with local electrostatic, the interaction energy between the charge and the surface can be calculated with the method of image charges.
	
	\begin{figure}
		\includegraphics[scale=1.1]{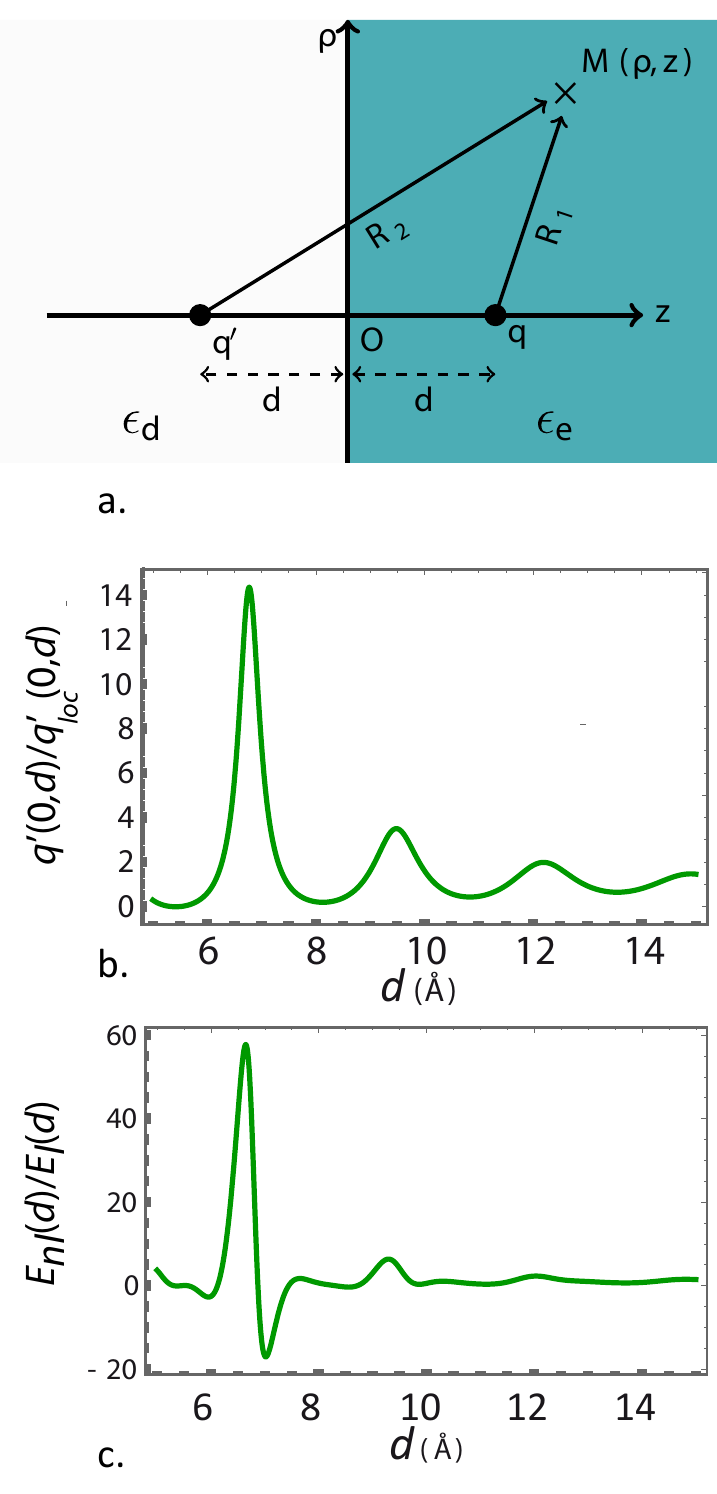}
		\caption{\textbf{a.} Schematic representation of a charge $q$ in a nonlocal medium interacting with a surface. The charge $q'$ is the image charge in the local medium.
			\textbf{b.} Image charge $q'$ as a function of the distance $d$. The expression given in Eq. (\ref{imcha}) is plotted for the parameter values given in Fig.2. and $\epsilon_d=1$ \textbf{c.} Interaction energy between a charge solvated in water and located at a distance $d$ of a local dielectric medium. The expression is given in Eq. (\ref{Eint}).}
	\end{figure}
	
	In the nonlocal medium, the electrostatic potential $\phi_{z>0}(\rho,z)$ at the point M $(\rho,z)$, can be written as the sum of a potential created by the real charge $q$ and by an image charge $q'$, located in $(0,0,-d)$,
	\begin{eqnarray}
	\phi_{z>0}(\rho,z)&=&q \Phi (R_1)+q'\Phi(R_2)\nonumber\\
	\label{phinl}
	\end{eqnarray}
	where $R_1(\rho,z)=\sqrt{(d-z)^2+\rho ^2}$ and  $R_2(\rho,z)=\sqrt{(d+z)^2+\rho^2}$ are the distances between M and $q$, M and $q'$ respectively, as illustrated on Fig.5 {\bf a.}, and where $\Phi(r)$ is obtained using Eq. (\ref{phi})

	\begin{equation}
	\label{Phi}
	\Phi(r)=\frac{1}{4\pi \epsilon_0 \epsilon_w r}\left(1+\epsilon_w h_{nl}(r)\right),
	\end{equation}
	with 
	\begin{equation}
	h_{nl}(r)=\frac{e^{-r/\lambda_e}\lambda_e^3\lambda_o^3\left(2 \lambda_e\lambda_o \cos(r/\lambda_o)+(\lambda_o^2-\lambda_e^2)\sin(r/\lambda_o)\right)}{2\alpha(\lambda_e^2+\lambda_o^2)^2}.
	\end{equation}
	with $\lambda_e$, $\lambda_o$ given in Eq. (\ref{lambda}) and $\alpha$ in Eq. (\ref{param3}).

	The electrostatic potential $\phi_{z<0} (\rho,z)$ in the local medium is simply equal to
	\begin{equation}
	\phi_{z<0}(\rho,z)=\frac{q''}{4\pi \epsilon_0 \epsilon_d R_1(\rho,z)}, 
	\label{pl}
	\end{equation}
	with $q''$ an image charge located in $(0,0,d)$.
	
	The expressions of the image charges $q'$ and $q''$ can be obtained by writing the boundary conditions for the 
	electrostatic field $\bf E$ and electrostatic displacement field ${\bf D}$ in $z=0$,
	\begin{equation}
	E^{z<0}_i= E^{z>0}_i,(i=x,y), \quad D^{z<0}_z=D^{z>0}_z.
	\end{equation}
	In a non local dielectric medium, these conditions will involve integro-differential equations due to the relation between ${\bf D}$ and ${\bf E}$, ${\bf D}_{\bf r'}=\int d^3r' \epsilon_{r,r'} {\bf E}_{\bf r'}$. 	To circumvent this difficulty, these conditions can be rewritten by introducing the electrostatic potential $\phi$, ${\bf E}=-\nabla \phi$ and a potential $\psi$ such that  ${\bf D}=-\nabla \psi$. It gives 
	\begin{eqnarray}
	\label{boundphi}
	\partial_{\rho}\phi_{z<0}(z=0)&=&\partial_{\rho}\phi_{z>0}(z=0), \\
	\partial_z \psi_{z<0}(z=0)&=&\partial_z \psi_{z>0}(z=0).
	\label{boundpsi}
	\end{eqnarray}
	One sees easily that for two local media, ${\bf D}=\epsilon_d {\bf E}$ for $z<0$ and  ${\bf D}=\epsilon_w {\bf E}$ for $z>0$, Eqs. (\ref{boundphi}-\ref{boundpsi}) take the usual form for the Maxwell boundary conditions \cite{jackson1999classical}. 
	
	Using the expressions of the potentials $\phi_{z>0}(\rho,z)$ and $\phi_{z<0}(\rho,z)$ given in Eqs. (\ref{phinl}-\ref{pl}) and $\psi_{z>0}=\frac{q}{4\pi \epsilon_0 R_1(\rho,z)}+\frac{q'}{4\pi\epsilon_0 R_2(\rho,z)}$ and $\psi_{z>0}(\rho,z)=\epsilon_d \phi_{z>0}$, one can solve Eqs. (\ref{boundphi}-\ref{boundpsi}) and find 
	\begin{eqnarray}
	\label{imcha}
	q'(\rho, d)&=&q \frac{\epsilon_w -\epsilon_d \left(1 +\epsilon_w h_{nl}(r)\right)+\epsilon_w \epsilon_d r h_{nl}'(r)}{\epsilon_w +\epsilon_d \left(1 +\epsilon_w h_{nl}(r)\right)-\epsilon_w \epsilon_d r h_{nl}'(r)}\\
	q''(\rho, d)&=&q-q'(\rho, d)
	\end{eqnarray}
	with $r=\sqrt{d^2+\rho^2}$. For large $r$, the image charges $q'$ and $q''$ tend to the values obtained for two local media, $q'_{loc}=q(\epsilon_w-\epsilon_d)/(\epsilon_w+\epsilon_d)$ and $q''_{loc}=2 \epsilon_d/(\epsilon_w+\epsilon_d)$ as the nonlocal correction of the electrostatic potential $h_{nl}(r)$ vanishes for large $r$.  The ratio $q'(d)/q'_{loc}$ for a point M located on the $z$ axis, $(\rho=0)$, is plotted on Fig. 5 {\bf b.} as a function of the distance $d$.
	
	One sees that the charge $q'(d)$ is a function of $d$ at short distance, contrary to the local case for which the image charge $q'_{loc}$ is constant. It exhibits oscillations alternating smaller and larger values than  $q'_{loc}$.
	The interaction energy $E_{nl}(d)$ between the charge $q$ and the wall is thus equal to 
	\begin{equation}
	E_{nl}(d)=qq'(d)\Phi(2d)
	\label{Eint}
	\end{equation}
	with $q'(d)$ given in Eq. (\ref{imcha}) and $\Phi(r)$ the function given in Eq. (\ref{Phi}). 
	
	The nonlocal interaction energy $E_{nl}$  scaled by the interaction energy obtained in the local case $E_l(d)=\frac{qq'_{loc}}{8\pi \epsilon_0 \epsilon_w d}$ is plotted on Fig. 5 {\bf c.}. Whereas the local energy is monotonous, {\it i.e.} attractive in the case $\epsilon_d \ll \epsilon_w$  and repulsive for $\epsilon_d \ll \epsilon_w$, the energy presents here succession of local minima/ This complex interaction can be related to molecular simulations studying the water/air interfaces \cite{Jungwirth01}, also showing an oscillating structure, although ionic correlations seem to play a predominant role.
	
	\section{Conclusion}
	Estimating the electrostatic interactions in water for objects separated by nanometric distances is essential in the understanding of {\it in cellulo} biochemical processes \cite{koehl2009} or nanofluidic systems to name a few.
	
	The effective interaction between two ions in water can be evaluated using molecular dynamics simulations. The radial distribution function $g_{ij}(r)$ for species immersed in water, {\it i. e.} the probability to find a species $i$ at a distance $r$ of species $j$ obtained with MD simulations,  can then be inverted to obtain the effective interaction potential $U_{ij}(r)$. The inversion procedure can be realized {\it via} inverse Monte Carlo simulations \cite{lyubartsev1995}, Integral equation theory \cite{zerah1986,molina2009}, or Maximum entropy likelihood. However, these methods face the usual limits of the molecular dynamics simulations such as the limitation to a numerical expression of the potential and the time needed to collect enough statistics for the structure of the solution.
	
The nonlocal electrostatic functionals give rise to equations that are tractable analytically for simple geometries or numerically solvable for more complex systems. However,  when fitted on SPC/E water dielectric properties, they failed to reproduce the behaviour of bulk water, such as the dipolar correlation functions and consequently the interactions in the fluid \cite{fedorov2007,berthoumieux2015fluctuation}. 
	
	In this paper, we propose to parameterize the electrostatic functional of the polarization using the dielectric susceptibility of SPC/E water treated as a dipolar liquid. We show that this procedure gives much better results for reproducing the dipolar correlation function of water.
 In this framework, we derive analytically the interaction between systems of point charges and between a point charge and a dielectric surface. Charge-surface interaction is essential in nanoconfined electrolytes but is difficult to investigate in molecular dynamics simulations, as image charge effects are out-of-scope of standard simulations with rigid surfaces. Nonlocal electrostatics can be formulated to easily treat these cases, as illustrated in the last part of this work. The calculated force qualitatively reproduces the layering of the solvent characterized by well-defined minima with a period of 2 \AA. Nevertheless, the absence of hard core repulsion exacerbates the effect which in practice for simple low charge ions rarely exceeds thermal agitation. The next step will thus be able to take into account these effects coupling size and dielectric relaxation. The ultimate goal is a quantitative analytical theory of the interactions between charged solutes.

	\vspace{2mm}
	\textbf{Acknowledgements:} HB acknowledges support from the CNRS through the D\'efi INFINITI - AAP 2018.
	\section*{Appendix}

	\subsection{Procedure for MD simulations}
	In this work we used the DLpoly 4 MD software\cite{DLPOLYmanual}. Electrostatic interactions were treated with the Particle Mesh Ewald (PME) summation technique. For simulations of the bulk water solution we used 1000 SPC/E water molecules in a cubic box with periodic conditions. After a short equilibration run of 0.01 ns with $NVE$  ensemble, a long equilibration run of 2 ns was performed in $NPT$ using the Berensen thermostat $T=293.15$ K and pressostat ($P=10^{-3}$ katm). Finally a simulation of 16 ns was performed in $NVE$ ensemble and used to determine the radial distribution functions of pure water.  A 2 fs time step was used,  while electrostatic interactions were  computed  using  the  Ewald  summation  technique, in order to take into account the periodicity of the system.  The cutoff value for the Van der Waals interaction was kept slightly below the half-cell size not to encounter problems due to the fluctuations of the cell’s volume during the simulations occurring in $NTP$ ensemble.
	
	\subsection{Dielectric susceptibility of water as a dipolar fluid}
	In order to parameterize the model, we compute the dielectric susceptibility for the SPC/E water treated as a fluid composed of dipoles. To do so, we replace in the equilibrium configurations of SPC/E water obtained with MD simulations each molecule by a dipole composed of en effective positive charge $q_f$ located on the SPC/E Oxygen and a negative charge -$q_f$ located on the dipolar axis of the water molecule at a distance $d$ of the Oxygen, as represented in Fig. 1.
	We impose that the dipole moment of this charge distribution is equal to the dipole moment of the SPC/E water, i. e. $\mu$=2.2 D, which leads to $q_f d$=2.2 D.
	The charge distribution $\rho_c(r)$ of such a dipolar liquid can be written as
	\begin{equation}
	\rho_c(r)=\Sigma_{i=1}^N\left( -q_{f}\delta(r-r_{O,i})+q_{f}\delta(r-r_{Hf,i}\right)
	\end{equation}
	with $N$ the number of molecules $r_{O,i}$ the position of the Oxygen in the $i^{th}$ molecule, $r_{Hf,i}$ the position of the effective positive charge in the  $i^{th}$ molecule.
	Such a two charges fluid is named a {\it Dipolar Dumbell Model} and its longitudinal dielectric susceptibility has been extensively studied \cite{raineri92}. 
	The charge structure factor $S(q)$  is the sum of the intramolecular contribution $S^{(m)}(q)$ and the intermolecular contribution $S^{(d)}(q)$ and can be written as
	\begin{eqnarray}
	S_c(q)&=&S^{(m)}(q)+S^{(m)}(q), \quad {\rm with}  \\
	S^{(m)}(q)&=&\frac{2nq_f^2}{q^2}\left(1-\frac{\sin (q d)}{qd}\right),\\
	S^{(d)}(q)&=&\frac{nq_f^2}{q^2}\left(h_{OO}(q)+h_{HfHf}(q)-2h_{OHf}(q)\right)
	\end{eqnarray}
	where $n$=0.033 \AA$^{-3}$ is the density of water in normal conditions and $h_{ij}(q)$ is the Fourier transform of the pair correlation function\cite{hansen} $h_{ij}(r)=g_{ij}(r)-1$.   
	
	
	\bibliographystyle{unsrt}
	\bibliography{bibmarin2}
\end{document}